# Laboratory Experiments on the Radiation Astrochemistry of Water Ice Phases

Duncan V. Mifsud[1,2,†], Perry A. Hailey[1], Péter Herczku[2], Zoltán Juhász[2], Sándor T. S. Kovács[2], Béla Sulik[2], Sergio Ioppolo[3], Zuzana Kaňuchová[4], Robert W. McCullough[5], Béla Paripás[6], and Nigel J. Mason[1]

1. Centre for Astrophysics and Planetary Science, School of Physical Sciences, University of Kent, Canterbury CT2 7NH, United Kingdom

2. Atomic and Molecular Physics Laboratory, Institute for Nuclear Research (Atomki), Debrecen H-4026, Hungary

3. School of Electronic Engineering and Computer Science, Queen Mary University of London, London E1 4NS, United Kingdom

4. Astronomical Institute, Slovak Academy of Sciences, Tatranská Lomnica SK-059 60, Slovak Republic

5. Department of Physics and Astronomy, School of Mathematics and Physics, Queen's University Belfast, Belfast BT7 1NN, United Kingdom

6. Department of Physics, Faculty of Mechanical Engineering and Informatics, University of Miskolc, Miskolc H-3515, Hungary

† Corresponding author: duncanvmifsud@gmail.com

## ORCID Identification Numbers

| Name | ORCID |
|---|---|
| Duncan V. Mifsud | 0000-0002-0379-354X |
| Perry A. Hailey | 0000-0002-8121-9674 |
| Péter Herczku | 0000-0002-1046-1375 |
| Zoltán Juhász | 0000-0003-3612-0437 |
| Sándor T. S. Kovács | 0000-0001-5332-3901 |
| Béla Sulik | 0000-0001-8088-5766 |
| Sergio Ioppolo | 0000-0002-2271-1781 |
| Zuzana Kaňuchová | 0000-0001-8845-6202 |
| Robert W. McCullough | 0000-0002-4361-8201 |
| Béla Paripás | 0000-0003-1453-1606 |
| Nigel J. Mason | 0000-0002-4468-8324 |

**Abstract**

Water ($H_2O$) ice is ubiquitous component of the universe, having been detected in a variety of interstellar and Solar System environments where radiation plays an important role in its physico-chemical transformations. Although the radiation chemistry of $H_2O$ astrophysical ice analogues has been well studied, direct and systematic comparisons of different solid phases are scarce and are typically limited to just two phases. In this article, we describe the results of an in-depth study of the 2 keV electron irradiation of amorphous solid water (ASW), restrained amorphous ice (RAI) and the cubic (Ic) and hexagonal (Ih) crystalline phases at 20 K so as to further uncover any potential dependence of the radiation physics and chemistry on the solid phase of the ice. Mid-infrared spectroscopic analysis of the four investigated $H_2O$ ice phases revealed that electron irradiation of the RAI, Ic, and Ih phases resulted in their amorphization (with the latter undergoing the process more slowly) while ASW underwent compaction. The abundance of hydrogen peroxide ($H_2O_2$) produced as a result of the irradiation was also found to vary between phases, with yields being highest in irradiated ASW. This observation is the cumulative result of several factors including the increased porosity and quantity of lattice defects in ASW, as well as its less extensive hydrogen-bonding network. Our results have astrophysical implications, particularly with regards to $H_2O$-rich icy interstellar and Solar System bodies exposed to both radiation fields and temperature gradients.

*Keywords:*   astrochemistry, planetary science, water ice, phase chemistry, electron-induced chemistry

**Graphical Abstract**

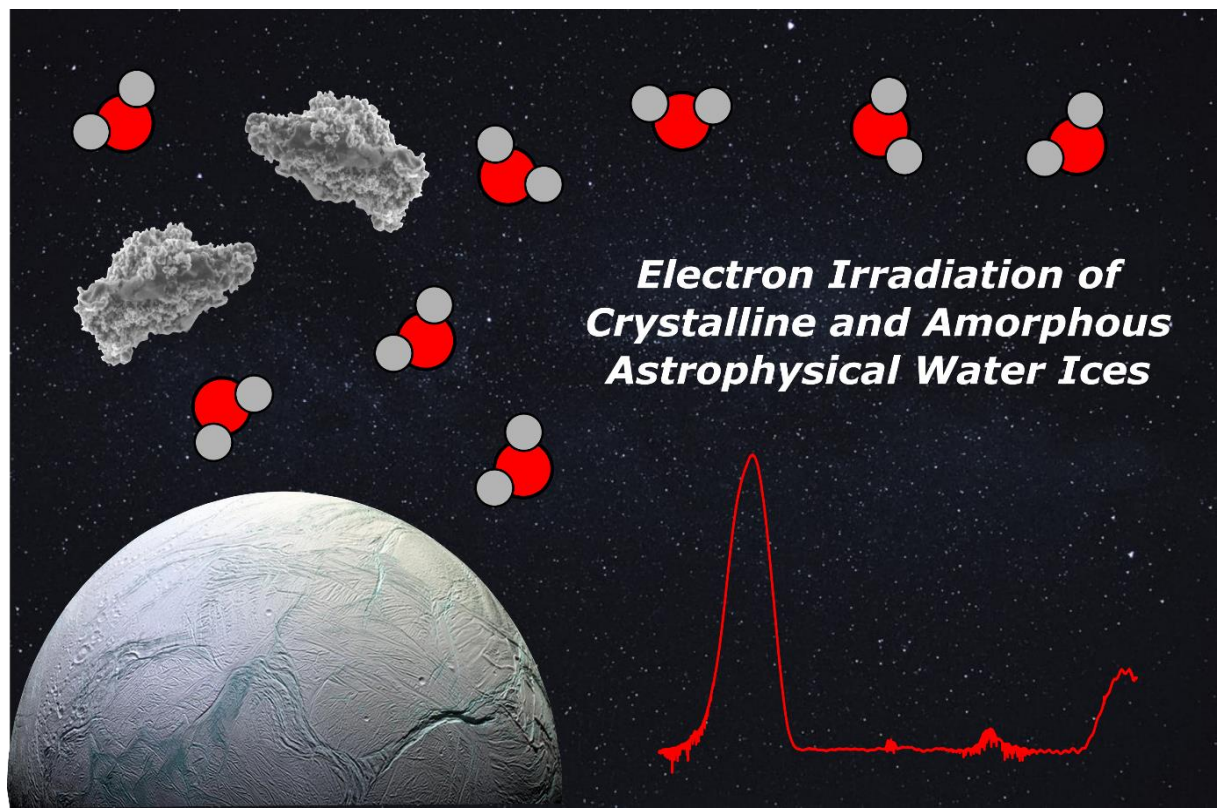

# 1  Introduction

Water ($H_2O$) ice is a ubiquitous material throughout the cosmos and has been the subject of several intensive research studies in planetary and space science [1-3]. On planet Earth, $H_2O$ ice largely exists as snow, glaciers, or permafrost, and their environmental transformations have been linked to various planetary scale climatological effects, such as altered ocean currents, changes in sea-level, and increased carbon exchange [4]. Beyond Earth, $H_2O$ ice has been detected in a great number of astrophysical settings, including dense interstellar clouds [5], circumstellar discs [6], comets [7,8], outer Solar System moons [9,10], rocky Solar System planets [11,12], and Kuiper Belt Objects [13].

Despite the low temperatures and ultra-high vacuum conditions associated with such space environments, astrophysical $H_2O$ ices display a rich phase chemistry (Fig. 1). Condensation of gas-phase $H_2O$ at the 10-20 K temperatures encountered within the cores of dense interstellar clouds results in the formation of amorphous solid water (ASW); a polymorphic and meta-stable phase whose crystallisation to a more ordered structure occurs spontaneously over extended time-scales [14]. The crystallisation of ASW can also occur *via* thermal annealing, with two crystalline phases forming sequentially: thermal annealing of ASW under astrophysical conditions initially results in its crystallisation to a cubic ice (Ic) which, upon further warming, is irreversibly re-structured to a stable hexagonal ice (Ih).

As ASW is thermally annealed, it begins to crystallise on a local scale *via* the nucleation of small regions or zones, thus leaving the majority of the ice in an amorphous state. Although still technically an amorphous ice, this phase is regarded as somewhat distinct to ASW due to its higher degree of crystallinity and has therefore been termed the restrained amorphous ice (RAI) phase [15]. RAIs have been detected under various experimental temperature and pressure conditions, including at temperatures as low as 77 K [16]. Further annealing of the ice results in its crystallisation to Ic at temperatures of 120-140 K [17], although just prior to this there exists a glass transition to a viscous liquid which is thermodynamically distinct from liquid $H_2O$ [18]. Further warming of the Ic phase results in an irreversible re-organisation to the stable Ih phase.

Both the Ic and Ih crystal structures consist of tetrahedrally co-ordinated $H_2O$ molecules arranged as puckered six-membered rings. The distinction between these crystal structures lies in the stacking order of the hexagonal bilayers which form the lattice, resulting in the Ic and Ih crystals adopting *I4$_1$md* and *Cmc2$_1$* space groups, respectively [19-21]. An additional $H_2O$ ice phase which is relevant to astrophysics is that of hyper-quenched glassy water (HGW), which is an amorphous phase formed as a result of the rapid cooling of liquid $H_2O$ and is commonly seen at the surfaces of icy outer Solar System moons following meteoritic impacts [22].

$H_2O$ ices in astrophysical environments may be subjected to various types of ice processing. For instance, icy grain mantles in dense interstellar clouds may be subjected to processing by galactic cosmic rays and low wavelength ultraviolet photons (such as Lyman-α), while ices in the Solar System may undergo processing by the solar wind, magnetospheric plasmas, and higher wavelength ultraviolet and visible photons [23,24]. Thermal processing may also take place if the ices are proximal to a heat source, such as a nascent or evolved star [25].

Several laboratory studies have considered the physico-chemical effects of such processing by making use of $H_2O$ astrophysical ice analogues. Irradiations using charged particles (i.e., ions or electrons) have been shown to result in the radiolysis of the $H_2O$ molecular ice and the formation of radiation products. Proton irradiation was demonstrated to result in the production of molecular hydrogen and oxygen ($H_2$ and $O_2$), as well as hydrogen peroxide ($H_2O_2$); as has irradiation using heavier ions such as helium and argon [26-28]. Irradiation using electrons or vacuum ultraviolet photons results in a similar chemical productivity [29,30], while the implantation of reactive ions into $H_2O$ ice can engender the formation of new species which incorporate the projectile ion [31-33].

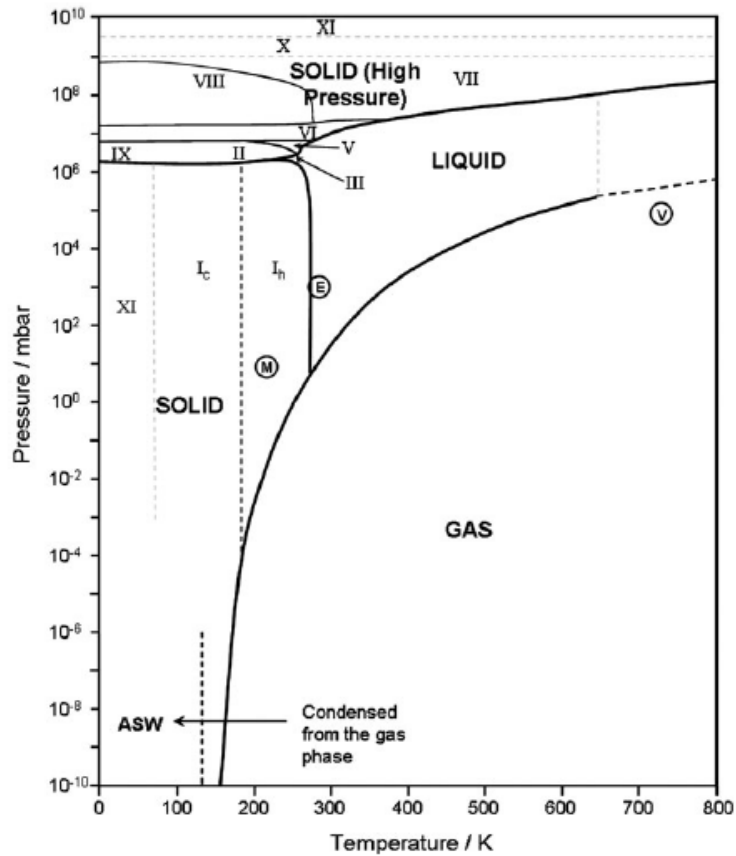

**Fig. 1** Phase diagram for $H_2O$ including pressure and temperature conditions relevant to astrophysics. Solid lines indicate conditions under which gaseous, liquid, solid, and high-pressure solid $H_2O$ exist, while dashed lines indicate approximate transition temperatures between different phases (note that transition temperatures indicated in the diagram are only indicative, as these may vary somewhat from one experimental set-up to another). Ambient conditions for Earth (E), Mars (M), and Venus (V) are indicated on the diagram. Reproduced from Mason *et al.* [34] with permission from the Royal Society of Chemistry.

The effects of radiation on the structure of $H_2O$ ices have also been investigated. Ion irradiation of crystalline $H_2O$ ices results in a temperature-dependent amorphization of the ice, with greater degrees of amorphousness being exhibited after irradiation at lower temperatures [35-38]. Such $H_2O$ ice amorphization is not unique to ion irradiation, and has also been documented after irradiation using electrons or ultraviolet photons [39-41]. In contrast, selective pumping of the $H_2O$ mid-infrared vibrational modes can result in the crystallisation of ASW [42]. Aside from such alterations in the phase of the $H_2O$ ice, radiation may also induce changes in its physical characteristics: ASW deposited from the gas phase is microporous, meaning both radiation products and adsorbates may be trapped within the pores and thereby influence the chemical pathways followed. Upon irradiation by charged particles, however, such microporosity decreases as the ice undergoes a process of compaction [43].

An in-depth understanding of the radiation physics and chemistry of astrophysical $H_2O$ ices is key to answering several fundamental questions in planetary and space science, including those regarding the potential habitability of icy outer Solar System worlds, as well as the chemical productivities of interstellar icy grain mantles. Although the radiation chemistry of $H_2O$ astrophysical ice analogues has been well probed in the laboratory using a variety of analytical techniques, there have been considerably fewer studies systematically comparing the irradiations of ASW and its crystalline phases.

Grieves and Orlando investigated the 0.1 keV electron irradiation of an unspecified crystalline phase of deuterated water ($D_2O$) as well as its microporous amorphous phase at 55 K [44]. They found that the chemical productivity of the amorphous ice was greater than that of the crystalline ice, with increased abundances of molecular deuterium ($D_2$) and $O_2$ being recorded when highly porous ASW was irradiated. This result was attributed to the increased diffusion of the relevant radicals and ions through pores and along grain boundaries, which are present to a greater extent in amorphous ices. Follow-up studies by Zheng *et al.* compared the 5 keV electron irradiation of the ASW and Ic phases of $D_2O$ at 12 K and reported similar results, but also invoked the possibility of increased L-type Bjerrum defects in ASW as a cause for the higher abundances of deuterium peroxide ($D_2O_2$) observed after irradiation of that phase compared to Ic [45].

Our recent experimental work studying the comparative 2 keV electron irradiation of amorphous and crystalline methanol ($CH_3OH$) and nitrous oxide ($N_2O$) ices at 20 K has also revealed a greater chemical productivity in irradiated amorphous solids [46]. However, that work additionally demonstrated the role of hydrogen-bonding in controlling the radiolytic decay of a molecular solid: crystalline solids which exhibit extensive arrays of hydrogen-bonds are more resistant to radiolytic decay than are their amorphous counterparts, thus providing fewer radicals for the formation of new products.

In this present investigation, we have performed comparative and systematic 2 keV electron irradiations of ASW, RAI, Ic, and Ih $H_2O$ ices at 20 K. Constraining the influence of the phase of an ice on its radiation chemistry is an important and oftentimes neglected aspect of experimental astrochemistry. Indeed, to the best of our knowledge, no previous study has included the RAI and Ih phases in such a comparative and systematic investigation. Our work therefore represents a progression in our understanding of the phase-dependence of the radiation physics and chemistry of $H_2O$ astrophysical ices, thus allowing this parameter to be better incorporated into more exhaustive and relevant future experiments performed using a systems astrochemistry approach [47]. We have also interpreted our results in light of the new conclusions presented by Mifsud *et al.* with regards to the role of hydrogen-bonds in impeding the radiolytic decay of crystalline ices [46].

## 2. Methods and Materials

Comparative electron irradiations of ASW, Ic, and Ih were performed using the Ice Chamber for Astrophysics-Astrochemistry (ICA) located at the Institute for Nuclear Research (Atomki) in Debrecen, Hungary. The experimental set-up has been described in detail previously [48,49], and so we limit ourselves to only a brief description of the most salient details here. The ICA is a high-vacuum chamber with a base pressure of a few $10^{-9}$ mbar which is maintained *via* the combined use of a dry rough vacuum pump and a turbomolecular pump. Within the centre of the chamber is a vertically-mounted gold-coated copper sample holder which carries a series of zinc selenide (ZnSe) substrates. The sample holder and substrates may be cooled to 20 K using a closed-cycle helium cryostat, although in principle a temperature control system allows for an operational temperature range of 20-300 K.

H$_2$O astrophysical ice analogues were prepared by background deposition of the vapour onto the cooled ZnSe substrates. De-ionised liquid H$_2$O was first de-gassed in a glass vial using the standard freeze-thaw-pump technique, after which vapour from the liquid was directed into the main chamber through an all-metal needle valve to allow for ice deposition at a pressure of a few $10^{-6}$ mbar. The deposition could be followed *in situ* using Fourier-transform mid-infrared transmission absorption spectroscopy (spectral range = 4000-650 cm$^{-1}$, resolution = 1 cm$^{-1}$). Different solid phases of H$_2$O were obtained by deposition at specific temperatures.

In order to assess the deposition temperatures required to synthesise the various phases of H$_2$O ice, a 0.25 µm thick ASW sample was prepared by deposition at 20 K and subsequently warmed up to 110 K at a rate of 1 K min$^{-1}$ with mid-infrared spectra collected at 10 K intervals, and thereafter at a slower rate of 0.33 K min$^{-1}$ with spectra being collected at 5 K intervals until sublimation was recorded at 165 K. Identification of the phase of H$_2$O ice present at a given temperature was achieved by comparison with previously published spectra [50-56].

Thermal annealing of the deposited ASW resulted in its crystallisation. As the ice was warmed, observed changes in the acquired mid-infrared spectra indicated a gradual structural transition to the Ic phase. Such spectroscopic indications could be seen at temperatures as low as 80 K, however this temperature is far too low to expect full crystallisation to Ic to have occurred and so we suggest that the observed spectroscopic signatures of crystallinity are more likely to be instead indicative of a RAI phase, in which pockets of crystalline ice are surrounded by a largely amorphous structure. Further annealing of the ice resulted in more extensive crystallisation, with the strongest spectroscopic indications of the Ic phase being observed over the 130-140 K temperature range. Beyond 140 K, spectroscopic changes indicated a morphological change to the Ih phase which was complete by the time a temperature of 145 K was reached. Continued warming of the ice resulted in sublimation-induced losses, and full desorption of the ice from the ZnSe deposition substrate occurred by 165 K. Fig. 2 summarises the main phases changes induced in the ice as a result of thermal annealing.

The thickness $d$ (µm) of the deposited ices was determined spectroscopically using Eq. 1 [48,49]:

$$d = 10^4 \frac{N_{H_2O} Z_{H_2O}}{N_A \rho} = 10^4 \left(\frac{P_{H_2O} \ln(10)}{A_v}\right)\left(\frac{Z_{H_2O}}{N_A \rho}\right)$$

(Eq. 1)

where $N$ is the column density of H$_2$O ice present (molecules cm$^{-2}$), $Z$ is the molecular mass of the H$_2$O molecule (18 g mol$^{-1}$), $N_A$ is the Avogadro constant (6.02×10$^{23}$ molecules mol$^{-1}$), and $\rho$ is the density of the ice, which we have taken to be 0.94 g cm$^{-3}$ for ASW and RAI, and 0.93 g cm$^{-3}$ for the Ic and Ih phases [29,57-59]. Note that the H$_2$O column density is measured spectroscopically by measuring the peak area $P$ (cm$^{-1}$) of a characteristic H$_2$O mid-infrared absorption band and dividing this by the band strength constant $A_v$ (cm molecule$^{-1}$) and subsequently multiplying the ratio by ln(10) [48]. In this study, we have measured the peak area of the prominent O–H stretching mode centred at about 3250 cm$^{-1}$ and have taken $A_v$ to be 2.0×10$^{-16}$ cm molecule$^{-1}$ for ASW and RAI, and 2.7×10$^{-16}$ cm molecule$^{-1}$ for the Ic and Ih phases [59,60].

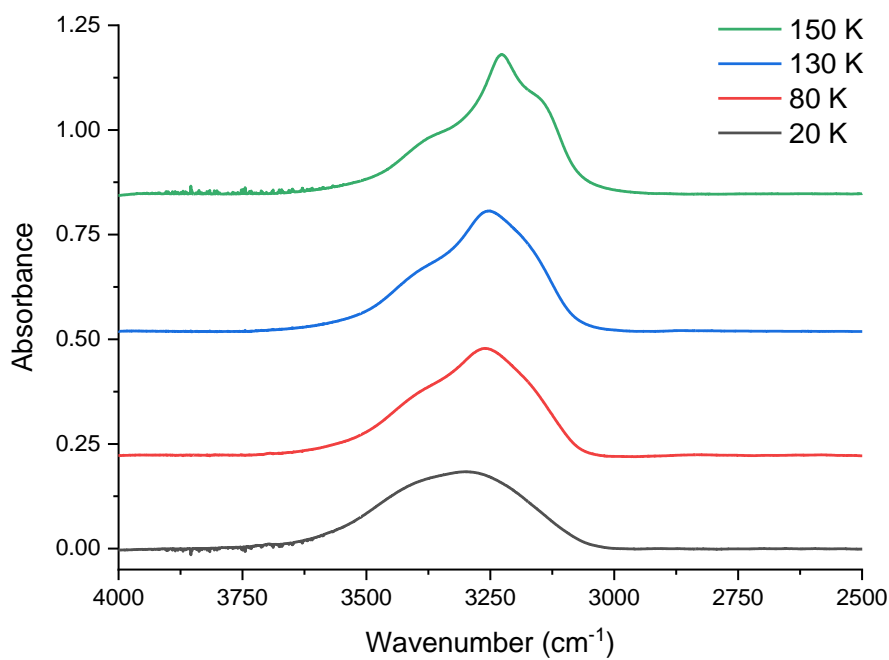

**Fig. 2** Evolution of the H$_2$O mid-infrared stretching mode as a result of thermal annealing of an ice from deposition at 20 K to sublimation at 165 K. Shown in this figure (from bottom top) are the appearances of this vibrational mode for the ice in the ASW (20 K), RAI (80 K), Ic (130 K), and Ih (150 K) phases. Note that spectra are vertically offset for clarity, with the scale of the $y$-axis being set to the lower most spectrum.

In order to constrain any potential phase dependence of the radiation chemistry of H$_2$O astrophysical ice analogues, a total of four ice phases were investigated: (i) one ASW formed by slow deposition of H$_2$O vapour at 20 K, (ii) one RAI formed by slow deposition at 80 K, (iii) one Ic formed by slow deposition at 130 K, and (iv) one Ih formed by slow deposition at 150 K. After deposition, each of the four ices was cooled to a pre-selected irradiation temperature of 20 K. The irradiation of all ices at the same temperature ensured that no differences in the observed radiation chemistry between the ice phases could be due to increased radical mobility in the ice matrix at higher temperatures [61,62].

Ices were then irradiated using a 2 keV electron beam with projectile electrons impacting the target ices at an angle of 36° to the normal. The electron beam current and homogeneity were measured prior to commencing irradiation using the method described by Mifsud *et al.* [49], and a total fluence of 1.3×10$^{17}$ electrons cm$^{-2}$ was delivered to each ice with mid-infrared spectra being collected at several intervals. CASINO simulations (details of which were given by Drouin *et al.*, [63]) showed that impinging 2 keV electrons penetrate to a depth of approximately 170-180 nm within the H$_2$O ices (Fig. 3). Given that the thicknesses of the irradiated ices were 0.21-0.23 μm, electrons were therefore effectively implanted into the ices. The irradiations performed in this study are summarised in Table 1.

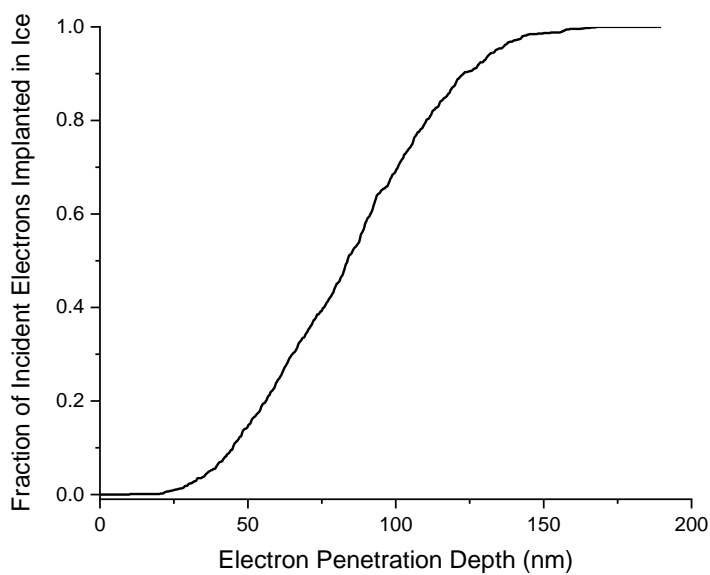

**Fig. 3** Results of a CASINO simulation showing the cumulative fraction of incident 2 keV electrons which are implanted into the ice when impacting ASW or RAI ($\rho$ = 0.94 g cm$^{-3}$) at an angle of 36° to the normal as a function of penetration depth. Results from similar simulations concerning Ic and Ih ice phases ($\rho$ = 0.93 g cm$^{-3}$) were very similar and are omitted from this figure for clarity.

**Table 1** Characteristics of the H$_2$O astrophysical ice analogues investigated in this study.

|  | **ASW** | **RAI** | **Ic** | **Ih** |
|---|---|---|---|---|
| Deposition Temperature (K) | 20 | 80 | 130 | 150 |
| Ice Thickness (μm) | 0.23 | 0.21 | 0.21 | 0.22 |
| H$_2$O Column Density (molecules cm$^{-2}$) | 7.4×10$^{17}$ | 6.5×10$^{17}$ | 6.6×10$^{17}$ | 6.8×10$^{17}$ |
| Ice Density (g cm$^{-3}$) | 0.94 | 0.94 | 0.93 | 0.93 |
| Irradiation Temperature (K) | 20 | 20 | 20 | 20 |
| Incident Electron Energy (keV) | 2 | 2 | 2 | 2 |
| Total Electron Fluence (electrons cm$^{-2}$) | 1.3×10$^{17}$ | 1.3×10$^{17}$ | 1.3×10$^{17}$ | 1.3×10$^{17}$ |
| Maximum Electron Penetration Depth (nm) | 178 | 178 | 168 | 168 |

## 3. Results

### 3.1 Characteristics of Mid-Infrared Spectra

The mid-infrared spectra of the various H$_2$O ice phases investigated in this study, both before and after irradiation using 2 keV electrons, are shown in Fig. 4. In the pre-irradiative spectra, several of the vibrational modes of the H$_2$O molecule are visible, including the bending mode ($v_2$) centred about 1655 cm$^{-1}$ and the libration mode ($v_L$) observed at 764 cm$^{-1}$ in ASW and at 819 cm$^{-1}$ in RAI, Ic, and Ih [29,45,64]. A weaker absorption band at 2206 cm$^{-1}$ is attributed to either the 3$v_L$ overtone band or the $v_L + v_2$ combination band. The broad absorption feature centred at about 3250 cm$^{-1}$ is a composite band formed of individual fundamental vibrational stretching modes, namely the in-phase and out-of-phase symmetric stretching ($v_1$) and the transverse and longitudinal asymmetric stretching modes ($v_3$).

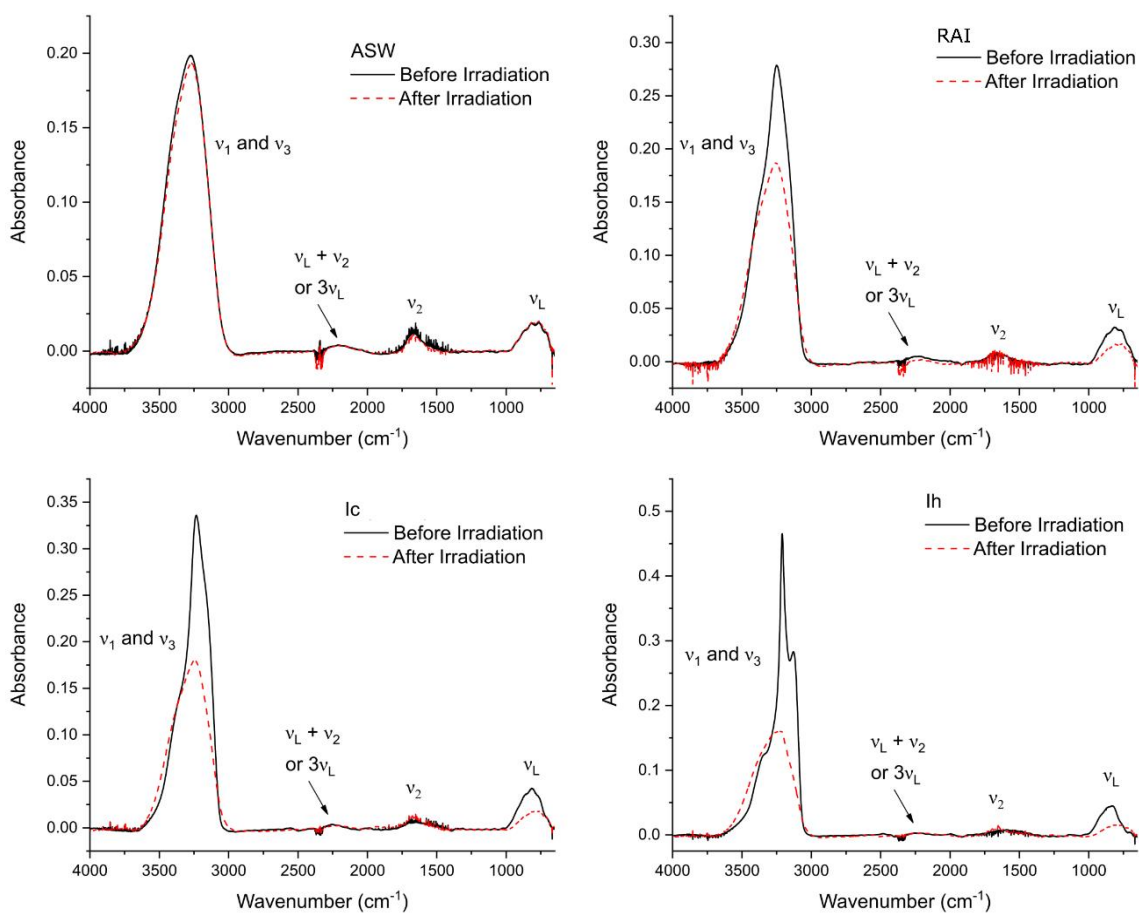

**Fig. 4** Mid-infrared spectra of ASW (deposited at 20 K), RAI (deposited at 80 K), Ic (deposited at 130 K), and Ih (deposited at 150 K) $H_2O$ ice phases before and after irradiation using a fluence of $1.3 \times 10^{17}$ electrons cm$^{-2}$. Observed vibrational modes have been labelled.

The shape of this broad stretching mode changes on going from ASW to RAI to Ic to Ih (Figs. 2 and 4). In ASW, the band appears as a very broad, single-peaked structure centred at about 3272 cm$^{-1}$. On transitioning to RAI, the band adopts a narrower profile due to the presence of crystallites within the structure, although additional features are difficult to detect. In the Ic phase, however, shoulder peaks are visible at about 3152 cm$^{-1}$ and 3399 cm$^{-1}$ which we have ascribed to increased resolution of the $\nu_3$ and $\nu_1$ modes, respectively [29,45,64]. In the mid-infrared spectrum of Ih, these contributions are still better resolved and the band instead appears as a multi-peaked structure.

The onset of electron irradiation results in noticeable changes in the appearance of the mid-infrared spectra of the $H_2O$ ices (Fig. 4). Perhaps the most evident of these is the change in the appearance of the broad stretching mode. When RAI and the crystalline ices (Ic and Ih) are irradiated, this band changes from a narrow, structured, and partially resolved one to one which is broader and essentially featureless; similar to that of ASW. This is indicative of an efficient radiation-induced amorphization process, as discussed previously. When directly comparing these amorphized phases, it is possible to note that the amorphization of RAI and Ic occur more rapidly than that of Ih as virtually all observed features associated with crystallinity in these ices are lost after a fluence of $1.4 \times 10^{16}$ electrons cm$^{-2}$ is delivered. In the case of Ih, however, some small and identifiable peaks within the broader band persist until a fluence of $4.4 \times 10^{16}$ electrons cm$^{-2}$, after which point the band adopts a broad and featureless structure similar to that of ASW. The progress of the amorphization process as observed from changes in the shape and structure of the broad stretching mode is depicted in Fig. 5.

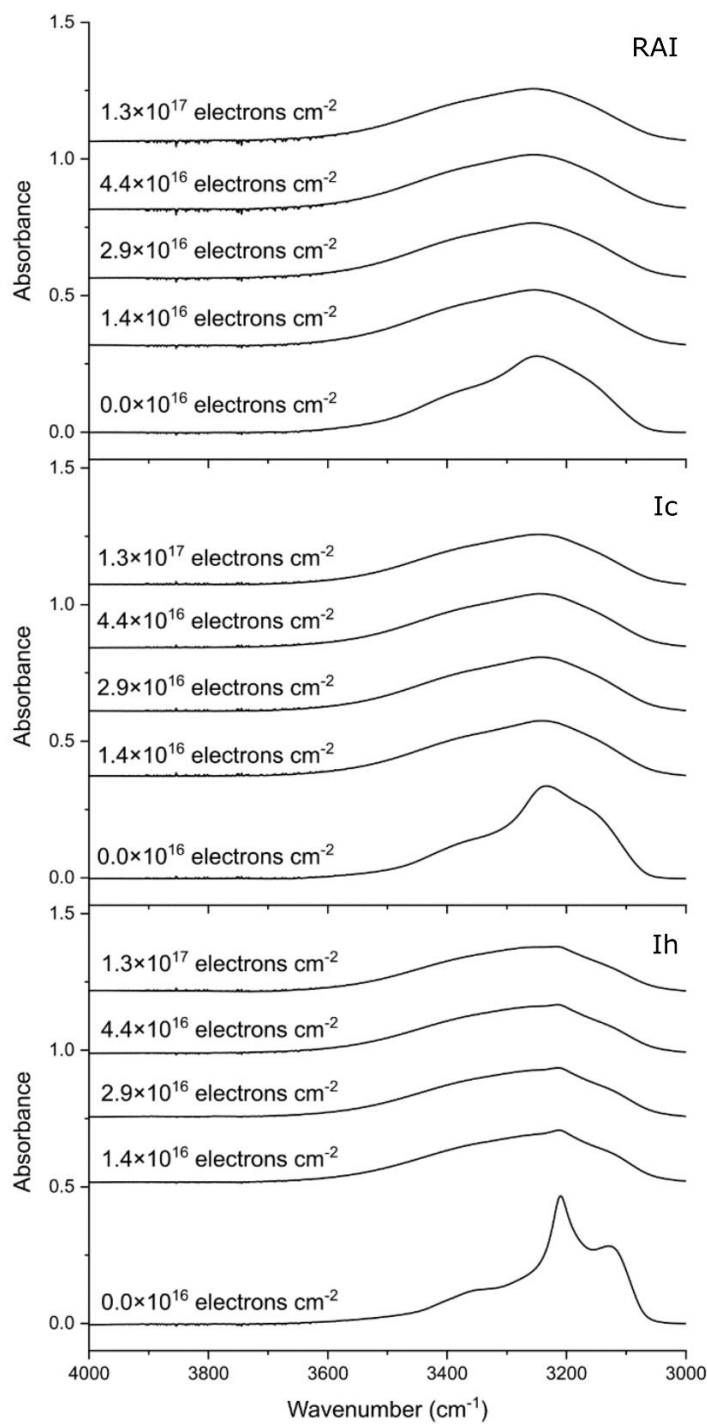

**Fig. 5** The structured H$_2$O bending mode visible in the mid-infrared spectra of Ih, Ic, and (to a lesser extent) RAI was observed to undergo increased broadening as a result of electron-induced amorphization of the ices. This amorphization is suggested to be most rapid in the RAI and Ic phases, for which all spectroscopic signatures of crystallinity were lost after an electron fluence of 1.4×10$^{16}$ electrons cm$^{-2}$ was supplied. The Ih phase seemed to be somewhat more resistant to radiation-induced amorphization, and crystallinity was only deemed to be lost after a fluence of 4.4×10$^{16}$ electrons cm$^{-2}$ was supplied. Note that spectra are vertically offset for clarity, with the scale of the y-axis in each panel being set to the lower most spectrum.

In contrast, the electron irradiation of ASW does not result in any significant change in the appearance of the broad stretching mode (Fig. 4). However, one change which was noted was the disappearance of two small peaks centred at 3722 cm$^{-1}$ and 3698 cm$^{-1}$ after the onset of irradiation (Fig. 6). These peaks are attributed to the hydroxyl (OH) dangling bonds in microporous ASW [65]. The disappearance of these peaks occurs early on during irradiation, with their complete absence being recorded at a fluence of $1.4\times10^{16}$ electrons cm$^{-2}$, and is attributed to a collapse of the ice pore structure leading to its compaction [43].

In addition to the physical re-structuring of the ice, electron irradiation also resulted in the formation of new chemical products. Although previous studies have demonstrated the formation of $H_2$ and $O_2$ as a result of the charged particle and ultraviolet photon irradiation of $H_2O$ astrophysical ice analogues [26-30,44,45,64], our mid-infrared spectroscopic analysis was unable to document the formation of these products due to their lack of a dipole moment which is necessary for analysis. The formation of $H_2O_2$, however, was observed through its characteristic combination band ($v_2 + v_6$) at approximately 2866 cm$^{-1}$ [66,67]. The symmetric bending ($v_2$) and asymmetric bending ($v_6$) modes were not observed. No spectroscopic evidence for the production of ozone ($O_3$) or the hydroperoxyl radical ($HO_2$) was observed, although this should not necessarily be interpreted as these species being absent from the ice as they may have been present but at concentrations below the spectroscopic limit of detection.

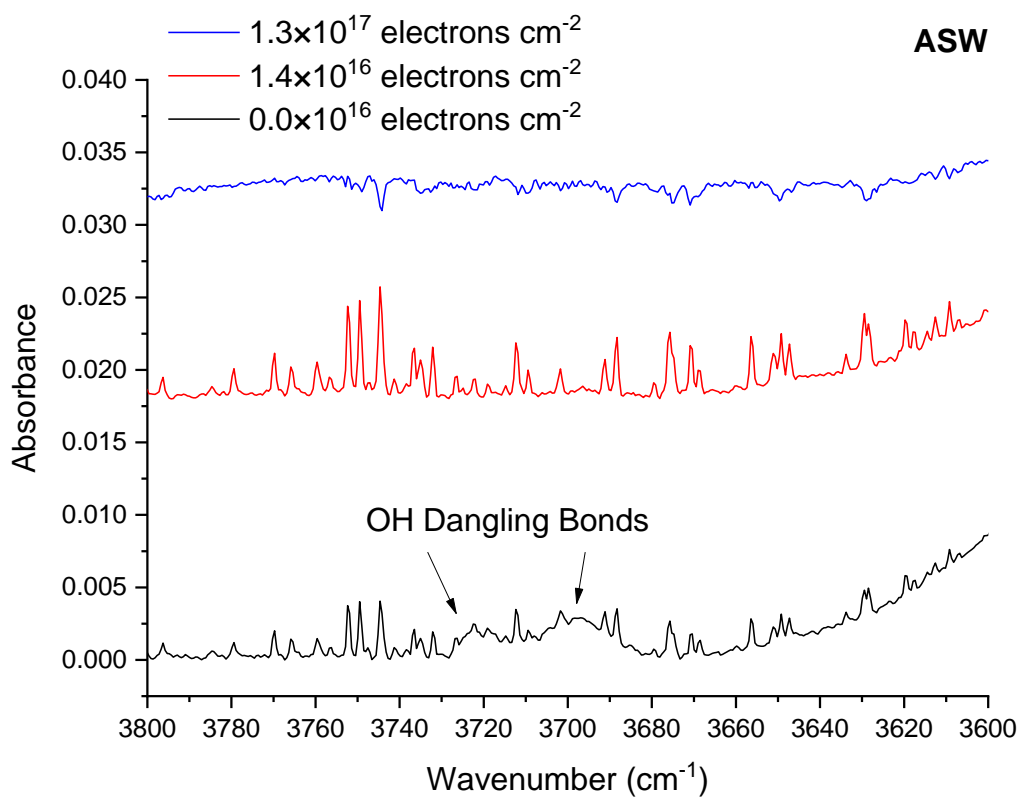

**Fig. 6** The OH dangling bonds in microporous ASW are visible as two small peaks (indicated by arrows) at the higher wavenumber end of the molecular stretching mode in the unirradiated ice (black trace; bottom). Once an electron fluence of $1.4\times10^{16}$ electrons cm$^{-2}$ has been supplied to the ice (red trace; middle), these peaks disappear indicating radiation-induced compaction of the ice and remain absent by the time a final fluence of $1.3\times10^{17}$ electrons cm$^{-2}$ has been delivered (blue trace; top). Note that spectra are vertically offset for clarity, with the scale of the *y*-axis being set to the lower most spectrum.

## 3.2 H$_2$O$_2$ Production

The abundance of H$_2$O$_2$ present in each H$_2$O astrophysical ice analogue after irradiation using 2 keV electrons was noted to differ between phases (Figs. 7 and 8). The production of H$_2$O$_2$ as a function of electron fluence was followed by measuring its column density *via* integration of the peak area of the combination band at 2866 cm$^{-1}$ (Eq. 1) and taking $A_v$ to be $5.7\times10^{-17}$ cm molecule$^{-1}$ [28]. H$_2$O$_2$ production was highest as a result of irradiation of ASW, while column densities observed after irradiation of the RAI phase were lower (Fig. 8). The electron irradiation of the Ic and Ih phases were similar to each other, but also resulted in the lowest abundances of H$_2$O$_2$ (Fig. 8). By comparing the initial column density of H$_2$O present before irradiation with the measured column density of H$_2$O$_2$ at the end of irradiation ($1.3\times10^{17}$ electrons cm$^{-2}$), it was possible to deduce that 0.32%, 0.21%, 0.16% and 0.14% of the H$_2$O was converted to H$_2$O$_2$ as a result of electron irradiation of ASW, RAI, Ic, and Ih, respectively.

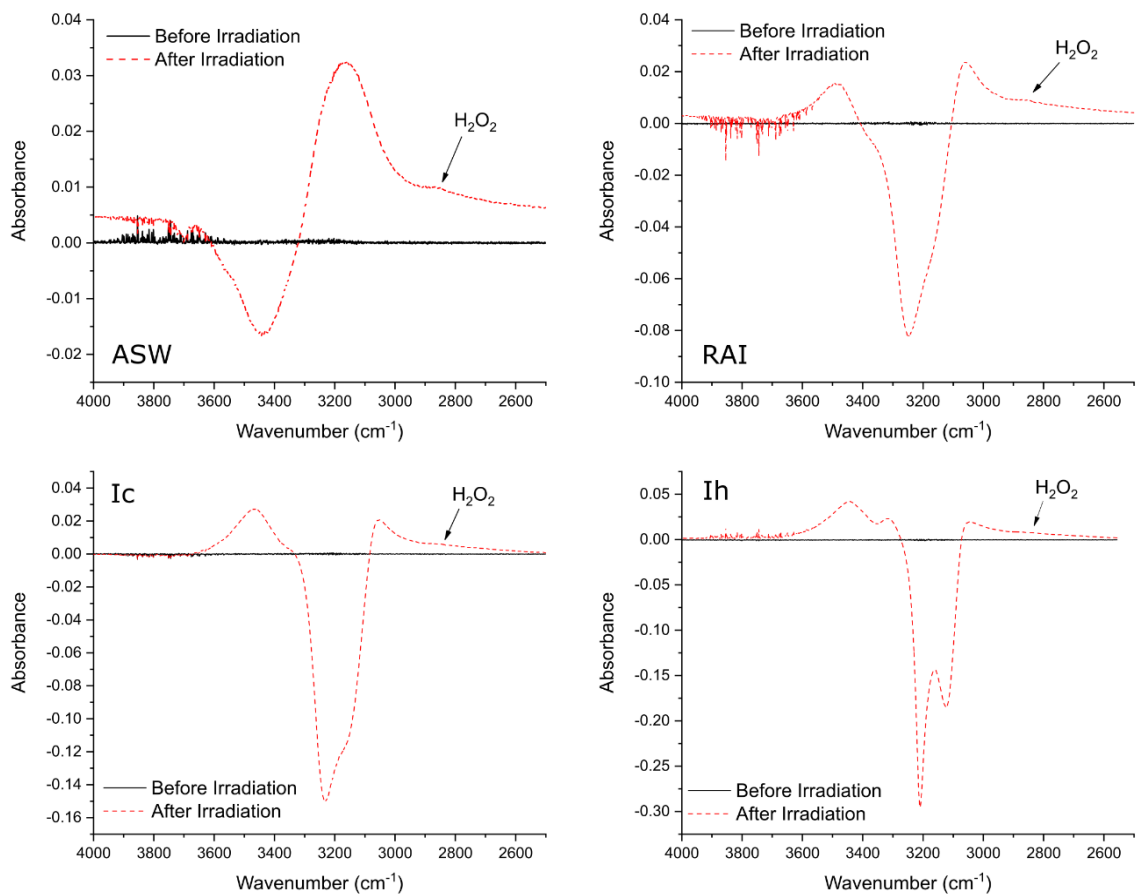

**Fig. 7** Spectral identification of H$_2$O$_2$ in electron irradiated ($1.3\times10^{17}$ electrons cm$^{-2}$) H$_2$O ices. In these spectra, the spectrum of the ice prior to any irradiation was used as a background, meaning radiolytic decay of the H$_2$O ice is seen as negative absorptions of its characteristic bands while the formation of product molecules is seen as more accentuated positive absorptions. Note that positive shifts associated with the H$_2$O stretching mode can also be seen in these spectra, and are due to a combination of changes in the physical structure of the ice (such as amorphization or compaction) which cause a change in the shape of the absorption band, as well as the appearance of the $v_5$ asymmetric stretching and $v_1$ symmetric stretching modes of H$_2$O$_2$ which are also in this region [68].

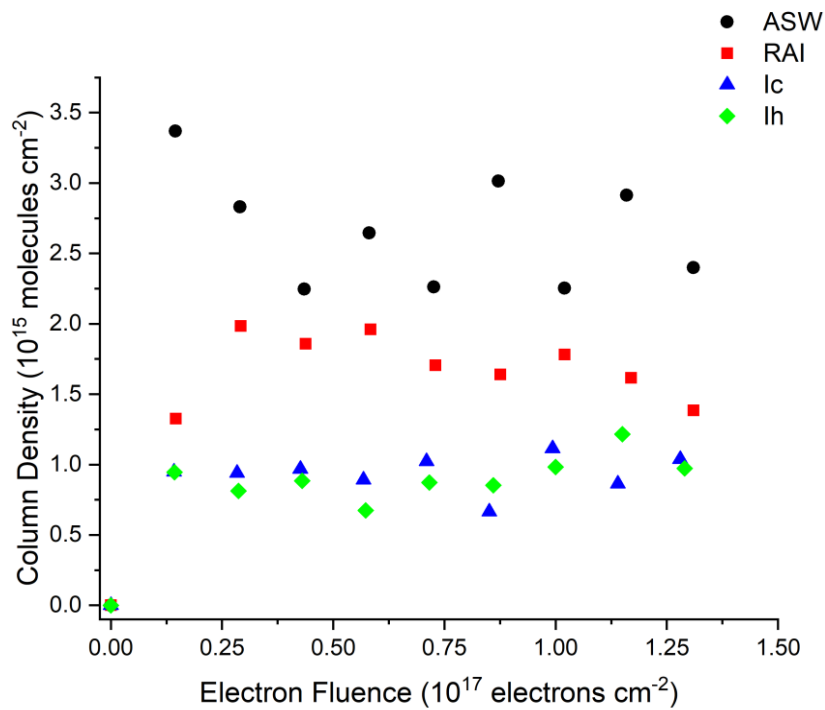

**Fig. 8** Scatter plot depicting the evolution of the measured $H_2O_2$ column density during 2 keV electron irradiation. Note that there is a progressive decrease in chemical productivity from ASW to RAI to Ic and Ih. Error bars have not been included, as the uncertainty in the column density is expected to be significantly less than the ~30% error associated with the band strength constant used to calculate it.

## 4. Discussion

### 4.1 Electron-Induced Structural Changes

The mid-infrared spectra depicted in Figs. 4 and 5 clearly show that electron irradiation of the $H_2O$ crystalline ice phases, as well as the RAI sample, results in their amorphization. This is seen through the broadening of various absorption features, but particularly in the stretching mode at approximately 3250 cm$^{-1}$ (Figs. 4 and 5). Several other studies have reported similar results [35,36,38,40,41]. The work of Leto and Baratta investigated the amorphization of Ic induced by Lyman-α photon irradiation and compared this with amorphization induced by 30 keV proton irradiation, and found that the latter was approximately twice as rapid [41].

Our results also allow for comparisons to be drawn; in this case between the amorphization of Ic and Ih. As stated previously, complete amorphization of the crystalline ice was observed spectroscopically after an electron fluence of 1.4×10$^{16}$ electrons cm$^{-2}$ was delivered to the Ic phase. However, the Ih phase was observed to be somewhat more resistant to amorphization, which was not fully observed spectroscopically until a fluence of 4.4×10$^{16}$ electrons cm$^{-2}$ had been delivered. This is not unexpected, as the Ih phase is a stable crystalline phase while the Ic phase is only meta-stable with respect to Ih [69]. This therefore suggests that the radiolytic phase transition from crystalline to amorphous, which is unachievable solely by cooling [14], is easier in the case of Ic.

The electron irradiation of Ic has also been studied in a series of papers by Zheng and co-workers [29,45,64], in which ~0.1 μm-thick ices were irradiated using 5 keV electrons to a final fluence of $3.6 \times 10^{17}$ electrons cm$^{-2}$. Although ice amorphization was reported in these studies, the process was not as complete as was observed in our results (Figs. 4 and 5), with several features of crystallinity still visible even at the end of their irradiation periods. The reason for this is likely relatively mundane: our electron beam was scanned over a much larger area of the ice meaning that there was a significantly lower proportion of unirradiated crystalline $H_2O$ ice to be observed spectroscopically in our data. In our experiments, the electron beam was scanned over an area which was >80% of the area scanned by the mid-infrared spectroscopic beam.

One interesting factor of crystalline $H_2O$ ice amorphization which has not been considered by the present study but which should nonetheless be discussed is the temperature dependence of this process. Earlier studies have revealed that the irradiation of crystalline $H_2O$ ice phases at lower temperatures leads to a greater extent of amorphization [35-37]. Indeed, the results of Zheng *et al.* [64], who investigated the temperature dependence of product formation as a result of the electron irradiation of Ic, attest to this, as the most extensive amorphization-induced changes in the Ic mid-infrared absorption bands were visible in the ice irradiated at 12 K, while no such changes could be observed after irradiation of Ic at 90 K. To the best of the authors' knowledge, amorphization of crystalline $H_2O$ ices as a result of electron irradiation has yet to be demonstrated at irradiation temperatures greater than 70 K. The reason for this is not fully understood, and constitutes a current avenue of investigation that our research group is pursuing.

Irradiation of ASW using 2 keV electrons did not result in any significant changes in the shape or appearance of its mid-infrared absorption bands (Fig. 4), other than minor band intensity decreases due to the radiolytic conversion of ASW to other molecules, such as $H_2O_2$. However, the two OH dangling bond peaks at 3722 cm$^{-1}$ and 3698 cm$^{-1}$ which are typically associated with microporosity were noted to disappear as a result of electron irradiation (Fig. 6). We have interpreted this observation as being the result of compaction of the ice. It is well known that ASW accreted from the vapour phase is a microporous structure, however the extent of this microporosity is dependent upon several parameters, including the temperature and rate of ice deposition [43].

Indeed, this becomes increasingly apparent when comparing our unirradiated ASW mid-infrared spectrum (Figs. 4 and 6) to that of Palumbo [43], who accreted a 0.25 μm thick ASW ice at 15 K from the vapour phase. Although the thicknesses of the ices are similar to our own, those of Palumbo display more prominent OH dangling bond peaks [43], indicating a greater extent of microporosity within the ice structure compared to ours. This is not unexpected, as previous studies have shown that the porosity of ASW is greater in ices accreted at lower temperatures [70,71]. It should also be noted that the angle of $H_2O$ deposition also influences the porosity of the condensed ice [72-74], however we do not believe that this caused the increased ice porosity observed in the study of Palumbo as both that study and this present one made use of a background deposition technique [43,48].

### 4.2 Comparative $H_2O_2$ Productivities

Our results appear to constitute the first comparative study of the electron-induced production of $H_2O_2$ from multiple solid phases of $H_2O$. The mechanistic chemistry occurring within the ice as a result of electron irradiation is complex and different radiolytic products are the result of following different chemical pathways [29,64,75]. Electron irradiation of a $H_2O$ molecule results in the radiolytic abstraction of either a hydrogen atom or an oxygen atom to yield OH or $H_2$, respectively (Eqs. 2 and 3). Abstracted hydrogen and oxygen atoms may combine to yield their respective homonuclear diatomic molecules, $H_2$ and $O_2$, or the heteronuclear OH radical (Eqs. 4-6).

$$H_2O \rightarrow H + OH$$

(Eq. 2)

$$H_2O \rightarrow H_2 + O$$

(Eq. 3)

$$H + H \rightarrow H_2$$

(Eq. 4)

$$O + O \rightarrow O_2$$

(Eq. 5)

$$O + H \rightarrow OH$$

(Eq. 6)

The production of $H_2O_2$ largely occurs as a result of addition reactions between these atoms and radicals. For instance, the combination of two OH radicals may produce $H_2O_2$ (Eq. 7). It is important to note, however, that at temperatures of about 10 K, where radicals are immobile within the bulk ice, this combination reaction can only occur if the OH radicals initially possess the correct geometry with respect to one another, where the distance between oxygen atoms in the separate radicals is minimised [29]. At temperatures higher than this, such as the 20 K irradiation temperature used in this present study, radicals are mobile and so the geometry of the combining OH radicals is likely to be less important.

The formation of $H_2O_2$ may also occur as a result of the reaction of an oxygen atom with a $H_2O$ molecule. Two barrierless reaction pathways are available [29,76]: the first is direct insertion of the electronically excited oxygen atom into the molecular structure of $H_2O$, yielding $H_2O_2$ (Eq. 8). The second route involves addition of the oxygen atom to the $H_2O$ molecule producing an oxywater intermediate ($H_2OO$), which then rearranges *via* hydrogen atom migration to form $H_2O_2$ (Eq. 9).

$$OH + OH \rightarrow H_2O_2$$

(Eq. 7)

$$O + H_2O \rightarrow H_2O_2$$

(Eq. 8)

$$O + H_2O \rightarrow H_2OO \rightarrow H_2O_2$$

(Eq. 9)

As can be seen in Fig. 8, the chemical productivity of $H_2O$ in terms of radiolytic $H_2O_2$ production was highest for the ASW phase, and proceeded to decrease by a factor of about two on moving to the RAI phase, and again by a factor of about two on moving to the Ic phase, which itself had a similar $H_2O_2$ productivity to the Ih phase. As such, it may be stated that there is a decrease in the production rate of $H_2O_2$ on transitioning from less ordered to more ordered ice phases. Similar trends for $O_2$ and $H_2$ production were observed by Zheng *et al.* who made use of quadrupole mass spectrometry to detect the formation of these products in electron irradiated ASW and Ic [45]. Such results are further evidence that the phase and structural morphology of an astrophysical ice analogue has a direct influence on its radiolytic chemical productivity [45,46].

We attribute the increased radiolytic $H_2O_2$ chemical productivity of less ordered $H_2O$ phases to a combination of at least three factors. Firstly, the microporous nature of ASW means that the OH radicals and hydrogen atoms formed as a result of radiolytic dissociation of the $H_2O$ molecule (Eq. 2) may migrate into the empty spaces left by the micropores [77]. Once there, hydrogen atoms may combine with one another to produce $H_2$ which may either desorb from the ice or remain trapped as a gas within the pores to form a clathrate-like structure. Either way, the reduction in the number of free hydrogen atoms renders the re-combination reaction to recycle $H_2O$ less probable, meaning there are more free OH radicals in the ASW phase available to produce $H_2O_2$ *via* Eq. 7. In a similar vein, the dissociation of $H_2O$ *via* Eq. 3 would also yield gas-phase $H_2$, as well as free oxygen atoms which may migrate along the surfaces of these pores and contribute to the increased production of $H_2O_2$ *via* Eqs. 8 and 9.

The second factor which may play a role in the greater abundance of $H_2O_2$ in electron irradiated ASW compared to the more ordered phases is the increased presence of structural defects in the former compared to the latter. Some of these defects, such as L-type Bjerrum defects, create conditions amenable to the production of $H_2O_2$. Bjerrum defects occur in $H_2O$ ices when, rather than a hydrogen-bonding network containing intermolecular interacting hydrogen and oxygen atoms being present, a $H_2O$ molecule is rotated about its oxygen atom such that two hydrogen atoms (D-type defect) or two oxygen atoms (L-type defect) face one another [78]. As such, L-type defects present ideal geometrical conditions for the direct linkage of two radiolytically produced OH radicals through their oxygen atoms so as to yield $H_2O_2$. The contribution of such defects to the production of $H_2O_2$ is thus likely to be greater at lower temperatures where OH radicals are immobile, as discussed previously.

Thirdly, it is likely that the presence of an extensive hydrogen-bonding network in crystalline $H_2O$ ice phases also contributes to their lower chemical productivities. In Ic and Ih, this hydrogen-bonding network extends throughout the entirety of the crystal lattice and thus adds an extra element of stability to the structure. During electron irradiation, a portion of the incident electron's kinetic energy must be expended upon overcoming the stabilising effect of hydrogen-bonding before radiolytic dissociation of the $H_2O$ molecule may take place [46]. Such a disruption would result in a localised amorphization characterised by increased ice defects (such as Bjerrum defects), as evidenced by the spectra presented in Fig. 5. In the ASW phase, hydrogen-bonding is significantly less extensive and so there is more energy to be imparted by the incident electron to drive radiolytic chemistry. However, the ASW phase derived from irradiated crystalline phases cannot be as chemically productive as the ASW phase accreted directly from vapour, since not only must a portion of the incident electron's kinetic energy be spent on disrupting the extensive hydrogen-bonding initially present, but the

ASW phase produced *via* radiolytic amorphization is likely to be devoid of micropores. Indeed, the mid-infrared spectra of electron irradiated Ic and Ih ices do not exhibit any peaks attributable to OH dangling bonds (Fig. 5).

Finally, we discuss the results of a previous study which investigated the temperature dependence of $H_2O_2$ production from electron irradiated Ic in the context of our own results [64]. In their investigation, Zheng *et al.* found that the abundance of $H_2O_2$ produced at lower irradiation temperatures was greater than that produced at higher ones, with a progressive decline in the amount of $H_2O_2$ observed being noted on going from 12 K to 90 K [64]. Similar trends were also noted in other studies investigating the ion irradiation of $H_2O$ ice at different temperatures [27,28,67]. Several explanations have been put forward to explain these results, including improved kinetics of the re-combination reaction of OH with a hydrogen atom to recycle $H_2O$ at elevated temperatures due to improved hydrogen atom diffusion coefficients [64], as well as increased $H_2O_2$ destruction rates *via* electron irradiation at higher temperatures [67].

Although we have not considered the effect of temperature in our study (indeed, our motivation was to study the variations in the observed radiation physics and chemistry of our ices induced by phase changes only), the results of those previous investigations into the temperature dependent formation of $H_2O_2$ together with the results presented in this study highlight the fact that simple changes in single experimental parameters may have remarkable changes in the chemistry of astrophysical ice analogues. However, fewer experiments have sought to investigate the effects of simultaneously changing multiple parameters on the resultant chemistry, despite this likely being more appropriate in terms of simulating astrophysical chemistry. Such an experimental approach is the basis of systems astrochemistry [47], and the data provided by this and similar studies will no doubt be useful in building up these more exhaustive and complete astrochemical experiments.

### 4.3 Astrophysical Implications

Our results carry important astrophysical implications since, as previously mentioned, $H_2O$ ice is ubiquitous in interstellar and Solar System environments. Although condensation of gas-phase $H_2O$ onto carbonaceous or silicate dust grains in the interstellar medium may result in the formation of microporous ASW, the majority of interstellar $H_2O$ ice is thought to be formed *via* surface atom addition reactions [79,80], and is thus likely formed as a compact ice. In any case, however, any microporous ASW deposited onto interstellar dust grains is expected to be compacted by impinging galactic cosmic rays within approximately $10^6$ years [59,81].

In contrast, in Solar System environments such as the Jovian and Saturnian satellites, $H_2O$ may condense from the gas phase after erupting from sub-surface oceans in plumes [82,83], leading to the formation of ice with increased porosity. Geophysical processes on these icy worlds may also influence $H_2O$ ice porosity [84]. These satellites also exist in radiation environments of their own which are mediated by the giant magnetospheres of their host planets, and so ASW present at the lunar surfaces may also undergo a process of compaction. Our results therefore contribute to and extend a growing body of research evidencing compaction of ASW as a result of irradiation using ultraviolet photons or charged particles [43,85-87].

Our results also demonstrate that the irradiation of crystalline $H_2O$ ice phases results in their amorphization, in agreement with previous studies [29,35-37,45,64]. Additionally, we have also demonstrated for the first time that the Ih phase is more resistant to radiation-induced amorphization than is the Ic phase by a fluence factor of about three. Indeed, $H_2O$ ices in most astrophysical environments are expected to undergo amorphization to some degree due to the radiation environments present. However, whether an astrophysical $H_2O$ ice is observed to be amorphous or crystalline is somewhat dependent upon the competing processes of radiation-induced amorphization and thermally-induced crystallisation. Icy interstellar dust mantles containing $H_2O$ in dense molecular clouds, for instance, may start out as ASW only to later crystallise due to increased temperatures during the birth and evolution of a nearby star [23].

In the Solar System, radiation processing by planetary magnetospheres or the solar wind would be expected to amorphize any $H_2O$ ices present on icy moons, comets, and Kuiper Belt Objects. However, these celestial bodies also experience warming and increased temperatures as a result of approaching perihelion during their orbit around the sun or of internal heating processes. Such warming could crystallise any ASW present on these bodies. Thus, an equilibrium $H_2O$ ice phase is generated which is largely amorphous when radiation processes dominate, and largely crystalline when thermal processes dominate. The three largest Galilean satellites of Jupiter provide an intriguing case study of this: $H_2O$ has been observed to be largely amorphous at the surface of Europa, where temperatures are low but magnetospheric ion fluxes are fairly high, but crystalline at the surface of Callisto where surface temperatures are higher [88]. On Ganymede, which possesses its own magnetosphere and where temperatures are intermediate between those at the surfaces of Europa and Callisto, ASW dominates at polar regions while crystalline phases are more common at equatorial regions which are shielded from impinging Jovian magnetospheric ions by the Ganymedean magnetosphere [89].

In terms of chemical productivity, our results have demonstrated that the electron irradiation of ASW results in a greater abundance of $H_2O_2$ than does the irradiation of crystalline phases, with a progressive decline in the amount of spectroscopically observed $H_2O_2$ on transitioning from ASW to RAI to Ic to Ih. Such a result therefore also implies that amorphous $H_2O$ astrophysical ices are also more chemically productive when subjected to processing by galactic cosmic rays, the solar wind, or magnetospheric ions compared to crystalline ones. The increased abundance and diversity of radiolytic product molecules in irradiated ASW may thus make it, and by extension those astrophysical environments where this phase is dominant, more conducive to the formation of other (potentially complex and prebiotic) molecules when other molecular species are also included in the ice.

## 5. Conclusions

In this study we have performed, for the first time, a comparative physico-chemical investigation of the 2 keV electron irradiation of the ASW, RAI, Ic, and Ih solid $H_2O$ phases under conditions relevant to astrophysics. Our results have demonstrated radiation-induced structural changes in the ice phases, including compaction of the ASW phase and amorphization of the more ordered phases. Amorphization of the Ic phase was noted to be more rapid than that of the Ih phase by a fluence factor of about three, likely due to the meta-stability of the former phase relative to the latter.

Radiation-induced chemical changes were also detected *via* mid-infrared spectroscopy, and the evolution of $H_2O_2$ could be monitored throughout irradiation. Our results have shown that the greatest abundance of $H_2O_2$ is formed after irradiation of ASW, with lower quantities formed on progressively transitioning to the more ordered and stable RAI, Ic, and Ih phases. Such results mirror similar findings documented after the 2 keV electron irradiation of other ices (including $CH_3OH$ and $N_2O$), and have been attributed to a combination of three factors: (i) the increased microporosity and (ii) structural defects which characterise the amorphous phase, thus allowing for a more efficient diffusion of radiolytically generated radicals away from their site of formation, and (iii) the more extensive hydrogen-bonding network in the crystalline phase which must be overcome before radiolytic chemistry can take place.

Finally, we note that the results of this study are relevant to our understanding of the chemistry of interstellar and Solar System environments. $H_2O$ ices are ubiquitous throughout the cosmos and the greater chemical productivity of irradiated ASW compared to its crystalline phases may imply that the chemistry leading to the formation of complex and potentially prebiotic molecules in ASW-dominated astrophysical environments is more extensive than in those settings where thermally-induced crystallisation of the $H_2O$ ice is a major process.

## Acknowledgements

The authors gratefully acknowledge funding from the Europlanet 2024 RI which has been funded by the European Union Horizon 2020 Research Innovation Programme under grant agreement No. 871149. The main components of the experimental apparatus used in this study were purchased using funding from the Royal Society through grants UF130409, RGF/EA/180306, and URF/R/191018.

Duncan V. Mifsud is the recipient of a University of Kent Vice-Chancellor's Research Scholarship. Sergio Ioppolo acknowledges the Royal Society for financial support. The research of Zuzana Kaňuchová is supported by VEGA – the Slovak Grant Agency for Science (grant No. 2/0059/22) and the Slovak Research and Development Agency (contract No. APVV-19-0072). The research of Béla Paripás is supported by the European Union and the State of Hungary and is co-financed by the European Regional Development Fund (grant GINOP-2.3.4-15-2016-00004).

## Author Contributions Statement

The experiment was designed by Duncan V. Mifsud and Perry A. Hailey and carried out by Duncan V. Mifsud, Péter Herczku, Zoltán Juhász, Sándor T. S. Kovács, and Béla Sulik. Data analysis was performed by Duncan V. Mifsud, who also wrote the manuscript. All authors took part in designing and building the experimental set-up, as well as in reviewing the manuscript.

## Conflict of Interests Statement

The authors hereby declare that this research was performed in the absence of commercial or financial relationships which may be construed as a conflict of interest.

## Data Availability Statement

Although the spectral data associated with this manuscript will not be uploaded to an online repository, interested readers are more than welcome to contact the corresponding author directly should they require more information regarding our original data files.